\documentstyle[12pt]{article}

\newenvironment{indention}[1]{\par%
\addtolength{\leftskip}{#1}%
\begingroup}{\endgroup \par}

\def\gev{{\hbox{GeV}}}

\begin{document}
\baselineskip 7.2mm

\begin{titlepage}

\begin{flushright}
DPNU-93-41 \\
AUE-06-93 \\
November \ 1993 \\
\end{flushright}

\vspace {1cm}

\begin{center}
{\Large{\bf A Large Majorana-Mass \\
                     From Calabi-Yau Superstring Models }}

\vspace {1cm}

Naoyuki HABA,${}^1$ \  Chuichiro HATTORI,${}^2$
\  Masahisa MATSUDA,${}^3$ \\
Takeo MATSUOKA${}^1$ and Daizo MOCHINAGA${}^1$ \\
\vspace {3mm}
${}^1${\it Department of Physics, Nagoya University \\
Nagoya, JAPAN 464-01} \\

${}^2${\it Department of Physics, Aichi Institute of Technology \\
Toyota, Aichi, JAPAN 470-03} \\

${}^3${\it Department of Physics and Astronomy \\
Aichi University of Education \\
Kariya, Aichi, JAPAN 448}

\end{center}

\vspace {1cm}

\begin{abstract}
In Calabi-Yau superstring models
it is found that two large intermediate energy scales
of symmetry breaking can be induced for special types of
the nonrenormalizable interactions.
In the models one set of $SO(10)$-singlet,
right-handed neutrino
and their mirror chiral superfields is needed.
Through the study of the minimization of the scalarpotential,
the conditions for the presence of two large intermediate
scales are obtained.
In this scheme a Majorana-mass possibly
amounts to $O(10^{9 \sim 10}\gev)$.
This large Majorana-mass solves the solar neutrino problem
and also is compatible with the cosmological bound
for stable light neutrinos.

\end{abstract}

\end{titlepage}

Superstring theory is the only known candidate of consistent
unification of all fundamental interactions.
Lacking a means of addressing the non-perturbative problems,
at present we are unable to select a true string vacuum
theoretically.
However, we can make use of phenomenological requirements on
superstring-derived models as a valuable clue to classify
the string vacua corresponding to a huge number of distinct
classical solutions.
{}From this point of view it is
important for us to understand how to connect
superstring theory with the standard model.

In Calabi-Yau superstring models the gauge symmetry
at the unification
scale is rank-6 or rank-5 and is larger than the standard gauge
symmmetry $G_{st}=SU(3)_c \times SU(2)_L \times U(1)_Y$ with rank-4
\cite{gaugesym}.
Consequently, there exist intermediate energy scales of symmetry
breaking between the unification scale and the electroweak scale.
Furthermore, in Calabi-Yau superstring models there appear extra
matter fields which are not contained in
the minimal supersymmetric standard model.
In fact, we generally have $G_{st}$-neutral but $E_6$-charged chiral
superfields and their mirror chiral superfields.
In order that Calabi-Yau superstring theory is brought into
contact with the standard model, some of these $G_{st}$-neutral
matter fields have to develop non-vanishing
vacuum expectation values(VEVs) at the
intermediate energy scales.

In the following we specialize in the case that the gauge symmetry at
the unification scale is rank-6.
In this case there should exist two intermediate energy scales
of the symmetry breaking,
which are represented by the VEVs
$\langle S \rangle $ and $\langle N \rangle $.
Here we denote an $SO(10)$-singlet chiral
superfield and a right-handed
neutrino chiral superfield $(\nu_R^c)$ as $S$ and $N$, respectively,
which belong to {\bf 27}-representation of $E_6$.
If these VEVs are sufficiently large compared with the soft
supersymmetry(susy) breaking scale $m_{SUSY}=O(10^3\gev)$,
we have to make the $D$-terms vanish at such large scales.
This is realized by setting
$\langle S \rangle = \langle \overline S \rangle $ and
$ \langle N \rangle = \langle \overline N \rangle $,
where $\overline S$ and $\overline N$
stand for mirror chiral superfields
of $S$ and $N$, respectively.
Since the superfield $S$ participates in a Yukawa interaction with
leptoquark chiral superfields, the order of magnitude of
$ \langle S \rangle (\langle \overline S \rangle )$
determines the lifetime of proton.
To be consistent with the proton stability, it is required that
$\langle S \rangle  \geq O(10^{16}\gev)$.
On the other hand, a non-vanishing $\langle N \rangle $
implies the lepton number violation.
Therefore, the magnitude of $\langle N \rangle
(\langle \overline N \rangle )$ seems to be closely
linked to a Majorana-mass(M-mass) of the right-handed neutrino.
Experimentally neutrino masses are so small compared with
quark masses and charged lepton masses.
Seesaw mechanism provides an interesting solution for the neutrino
mass problem by introducing large M-masses for
right-handed neutrinos
\cite{seesaw}.
If we take the solar neutrino problem seriously, the M-mass
of the right-handed neutrino should be
of order $10^{9 \sim 12}$\gev
\cite{neutrino}.
Also this large M-mass is compatible with the cosmological bound
for stable light neutrinos
\cite{Cowsik}.

As mentioned above, from the proton stability the condition
\begin{equation}
        \langle S \rangle  \geq O(10^{16}\gev)
\end{equation}
should be satisfied.
How can we derive such large intermediate scales in Calabi-Yau
superstring models?
The discrete symmetry of the compactified manifold possibly
accomplishes this desired situation
\cite{discrete}.
In superstring models there exist effective
non-renormalizable(NR) terms in the superpotential.
The order of magnitudes of $\langle S \rangle $ and
$\langle N \rangle $ are governed by these NR terms.
Along this fascinating line
the problems of two large intermediate scales of
symmetry breaking and mass matrices have been studied
first by Masip
\cite{Masip}.
In the analysis general structures of the scalarpotential
are not sufficiently clarified
and conditions on the NR terms for the presence of
two large intermediate scales are obscure.

In this paper we find the constraints on the NR terms for the
presence of two large intermediate scales
and of a large M-mass.
Furthermore, we show that two
intermediate scales of symmetry breaking are
\begin{equation}
        \langle S \rangle  \geq O(10^{16}\gev),
         \ \ \ \ \ O(10^{15}\gev) \geq
                         \langle N \rangle  \geq O(10^{13}\gev)
\end{equation}
for special types of the NR terms
and that a M-mass of right-handed neutrino becomes
$M_M \sim m_{SUSY} ( \langle S \rangle / \langle N \rangle )^2$.
Its numerical value of the M-mass
possibly amounts to $O(10^{9 \sim 10}\gev)$.

First we take up the NR interactions in the superpotential
coming from a pair of $S$ and $\overline S$ chiral superfields.
The NR terms are of the form
\begin{equation}
        W_{NR} = \sum_{p=2}^{\infty}
        \lambda_p M_C{}^{3-2p}(S\overline S)^p,
\end{equation}
where $M_C$ is the unification scale.
Dimensionless coupling $\lambda _p$'s are of order one.
However, if the compactified manifold has a specific type of
discrete symmetry, some of $\lambda _p$'s become vanishing.
For instance, in the four-generation model obtained from
the Calabi-Yau manifold with the high discrete symmetry
$S_5 \times Z_5{}^5$,
this symmetry requires that $\lambda _p=0$ for $p \neq 4$ (mod 5)
\cite{discrete}.
When we denote the lowest number of $p$ as $n$, the NR terms
are approximately written as
$ W_{NR} \cong \lambda_n M_C{}^{3-2n}(S\overline S)^n$.

To maintain susy down to a TeV scale,
the scalarpotential should satisfy $F$-flatness and $D$-flatness
conditions at the large intermediate scale.
Then we have to set
$\langle S \rangle =\langle \overline S \rangle $.
As far as $D$-terms are concerned, the VEV can be taken
as large as we want.
Incorporating the soft susy breaking terms,
we have the scalarpotential
\begin{eqnarray}
       V &=& n^2 \lambda _n{ }^2 M_C{ }^{6-4n}
             \left( \vert S\vert ^{2(n-1)}\vert
		\overline S\vert ^{2n}
             +\vert S\vert ^{2n}\vert \overline S
		\vert ^{2(n-1)} \right)
                                  \nonumber \\
         & & \mbox{ } + {\frac {1}{2}} \sum _{\alpha }
		g_{\alpha }{}^2
              \left( S^{\dag }T_{\alpha }S
                  - \overline S^{\dag }T_{\alpha }
		\overline S \right)^2
                       + V_{soft}, \\
       V_{soft} &=&  m_S{}^2\,\vert S\vert ^2
                       + m_{\overline S}{}^2\,
                       \vert {\overline S}\vert^2,
\end{eqnarray}
where the $T_{\alpha }$ are Lie algebra generators and
$m_S{}^2$ and $m_{\overline S}{}^2$ are
the running scalar masses squared from the soft susy breaking.
$S$ and $\overline S$ develop nonzero VEVs
when $m_S{}^2+m_{\overline S}{}^2  < 0$.
In the renormalization group analysis
for the four-generation model,
it has been proven that $m_S{}^2+m_{\overline S}{}^2$
possibly becomes
negative at the large intermediate scale $O(10^{16}\gev)$
\cite{Zoglin}.
By minimizing $V$, we obtain the VEVs as
\begin{equation}
       \langle S \rangle  \simeq \langle \overline S \rangle
                  \sim  M_C\left(
                  \frac {\sqrt {-m_S^2}}{M_C}\right)^{1/2(n-1)}.
\end{equation}
The difference $\langle S \rangle - \langle \overline S \rangle$
is negligibly small and
we put $m_{S}{}^2 = m_{\overline S}{}^2$ approximately.
In the case $n=4$ as in the four-generation model,
the intermediate energy scale becomes
\begin{equation}
       \langle S \rangle  \simeq \langle \overline S \rangle
                    \sim O(10^{16}\gev)
\end{equation}
for $M_C=10^{18\sim 19}$\gev.
If $n=2$, then we have $\langle S \rangle  \sim 10^{11}$\gev,
which leads to the fast proton decay.
Through the Higgs mechanism,
the $(S-\overline S)/\sqrt {2}$ are absorbed into a massive
vector superfield with its mass
of $O(g_{\alpha }\langle S \rangle )$.
The component $(S + \overline S)/\sqrt {2}$ have masses
of order $O(10^3\gev)$  irrespectively of $n$.
In the case of only a pair of $S$ and $\overline S$
it is impossible for us to get sufficiently large M-masses
compared with the soft susy breaking scale.

Next we proceed to study the NR terms
which consist of pairs of $S$, $N$ and
$\overline S$, $\overline N$ chiral superfields,
provided that there appear $S$, $N$ and
$\overline S$, $\overline N$ superfields
in adequate Calabi-Yau models.
In this case we potentially derive two intermediate
energy scales of symmetry breaking.
Here we assume the NR interactions
\begin{equation}
    W_{NR} = M_C{}^3 \,\biggl[\lambda _1\,
    			\frac {(S\overline S)^n}{M_C{}^{2n}}
                    + \lambda _2\,\frac
			{(N\overline N)^m}{M_C{}^{2m}}
                    + \lambda _3\,\frac
			{(S\overline S)^i(N\overline N)^j}
                                      {M_C{}^{2(i+j)}}\biggr],
\end{equation}
where $n,m,i$ and $j$ are generally integers with
\begin{equation}
          n,\ m  \geq 2, \ \ \ \ \ i,\ j \geq 1
\end{equation}
and $\lambda_i$'s are constants of $O(1)$.
In certain types of Calabi-Yau models we suppose that
the exponents $n, m, i$ and $j$ are settled on appropriate values
due to the discrete symmetry of the Calabi-Yau manifold.
By minimizing the scalarpotential including
the soft susy breaking terms
\begin{equation}
       V_{soft} =   m_S{}^2\,\vert S\vert ^2
                      + m_{\overline S}{}^2\,
			\vert {\overline S}\vert^2
                   + m_N{}^2\,\vert N\vert ^2
                      + m_{\overline N}{}^2\,
			\vert {\overline N}\vert^2,
\end{equation}
we can determine the energy scales of symmetry breaking,
that is, $\langle S \rangle$ and $\langle N \rangle$.
The scalar mass parameters $m_S{}^2$ and $m_N{}^2$
evolve according to the renormalization group equations.
As in the four-generation model, we expect that
$m_S{}^2$ becomes negative at the large intermediate scale($M_I$).
On the other hand, it is natural to expect that
$m_N{}^2$ remains still positive
at $M_I$ scale,
because $N$ and $\overline N$ have different gauge quantum numbers
from $S$ and $\overline S$ and also
have different Yukawa interactions.
Hereafter we take $m_S{}^2 < 0$ and $m_N{}^2 > 0$
at $M_I$ scale.
However, the sign of $m_N{}^2$ is not essential
in the following discussions.
{}From the $D$-flatness condition
we get $\langle S \rangle = \langle \overline S \rangle$
and $\langle N \rangle = \langle \overline N \rangle$
in the approximation $m_S{}^2 = m_{\overline S}{}^2$
and $m_N{}^2 = m_{\overline N}{}^2$.
To find solutions in which the VEVs are real,
here we parametrize as
\begin{equation}
       \langle S \rangle =
       \langle \overline S \rangle = M_C\, x, \ \ \ \ \
       \langle N \rangle =
       \langle \overline N \rangle = M_C\, y.
\end{equation}
For convenience' sake, instead of $\lambda _i$'s
we use the parameters $a,b$ and $c$ defined as
\begin{equation}
       \lambda _1 = \frac {a}{n}, \ \ \ \ \
       \lambda _2 = \frac {a}{m} \,b^{-2m}, \ \ \ \ \
       \lambda _3 = -\frac {ac}{ij} \,b^{-2j},
\end{equation}
where $b$ and $c$ are put as positive.
For negative $c$ we have no solutions in which
$x$ and $y$ are real.
Let us introduce two dimensionless functions $f$ and $g$ :
\begin{eqnarray}
      f(x,y) & \equiv &
          M_C{}^{-2} \left. \frac {\partial W}{\partial S}\right|
              = a \left(
                  x^{2n-1} -
                  \frac {c}{j} x^{2i-1}
			\left( \frac {y}{b}\right)^{2j}
                      \right), \\
      g(x,y) & \equiv &
          M_C{}^{-2} \left. \frac {\partial W}{\partial N}\right|
              = \frac {a}{b} \left(
                  \left(\frac {y}{b} \right)^{2m-1} -
                  \frac {c}{i} x^{2i}
			\left( \frac {y}{b} \right)^{2j-1}
                      \right),
\end{eqnarray}
where $\ldots \vert $ means the values at
$S = \overline S = \langle S \rangle$
and $N = \overline N = \langle N \rangle$.
By using the $D$-flatness condition we have the scalarpotential
\begin{equation}
       M_C{}^{-4}V\vert = 2\,f(x,y)^2 + 2\,g(x,y)^2
                          - 2\,\rho _x{}^2x^2 + 2\,\rho _y{}^2y^2
\end{equation}
with
\begin{equation}
        \rho _x{}^2 = - \frac {m_S{}^2}{M_C{}^2}\ (>0), \ \ \ \ \
        \rho _y{}^2 =   \frac {m_N{}^2}{M_C{}^2}\ (>0).
\end{equation}

We now turn to study the
absolute minimum of the scalarpotential $V$.
At the minimum point the conditions
\begin{equation}
      \frac {\partial V}{\partial S} =
       \frac {\partial V}{\partial \overline S} =
        \frac {\partial V}{\partial N} =
         \frac {\partial V}{\partial \overline N} = 0
\end{equation}
have to be satisfied.
In the present notation the conditions are expressed as
\begin {eqnarray}
        f\,f_x + g\,g_x - \rho _x{}^2\,x = 0, \\
        f\,f_y + g\,g_y + \rho _y{}^2\,y = 0,
\end{eqnarray}
where $f_x = {\partial f}/{\partial x}$ and so forth.
Solving the above equations
and calculating the second derivatives
such as $\partial ^2V / \partial S^2$,
we find local minima and saddle points.
Since the scalarpotential is symmetric under the reflection
$x \rightarrow -x$ and/or $y \rightarrow -y$,
it is sufficient for us to consider
only the first quadrant in the $x-y$ plane.

Let us consider the case $n/m = i/(m-j)$.
In this case at the region $x^n \sim y^m$
the first terms of Eqs. (13)
and (14) become the same order with
the second terms coincidentally.
This situation is of critical importance
in the minimization of the scalarpotential.
Furthermore, we take $n > m$ so that $\langle S \rangle
> \langle N \rangle$.
Thus we study the case
\begin{equation}
       \frac {n}{m} = \frac {i}{m-j} > 1.
\end{equation}
In this case it can be proven that
there are the following two (three) local minimum points
for $j=1$ $(j \geq 2)$.
The values of the scalarpotential at these points
are calculable.

\begin{indention}{5mm}
    {\bf Point \ A}: \ \ \ \ \ $(x,y)  =  (x_0,y_0)$.
\begin{equation}
          M_C{}^{-4}V  \cong
                - \frac {4(n-1)}{(2n-1)} \,\rho _x{}^2 x_0{}^2,
\end{equation}
\begin{indention}{5mm}
where
\begin{eqnarray}
            & x_0 &= \left(
                      \frac {\rho _x}{\sqrt {2n-1} \,a \,\xi }
                    \right) ^{1/2(n-1)} , \nonumber \\
            & y_0 &= b \left( \frac {c}{i} \right) ^{1/2(m-j)}
                       x_0{}^{i/(m-j)} \ \ \ (\ll x_0), \\
            & \xi &= \left| 1-\frac{i}{j} \left( \frac {c}{i}
                            \right) ^{n/i} \right|. \nonumber
\end{eqnarray}
\end{indention}
    {\bf Point \ B}: \ \ \ \ \ $(x,y)  =  (x_0',0)$.
\begin{equation}
          M_C{}^{-4}V  \cong
                - \frac {4(n-1)}{(2n-1)} \rho _x{}^2 {x_0'}^2,
\end{equation}
\begin{indention}{5mm}
where
\begin{equation}
             x_0' = \left(
                      \frac {\rho _x}{\sqrt {2n-1} \,a }
                    \right) ^{1/2(n-1)}.
\end{equation}
\end{indention}

    {\bf Point \ C}: \ \ \ \ \ $(x,y)  =
    (x_0',y_0')$ \ \ \ $(j \geq 2)$.
\begin{equation}
          M_C{}^{-4}V  \cong
                - \frac {4(n-1)}{(2n-1)} \,\rho _x{}^2 {x_0'}^2,
\end{equation}
\begin{indention}{5mm}
where
\begin{eqnarray}
            & y_0' &= b \left(
                         \frac {i^2b^2}{(2j-1)c}
                            \left( 1 + \sqrt {1+R} \right)
                       \right)^{1/2(j-1)}
                       {x_0'}^{(n-i-1)/(j-1)}
                               \ \ \ (\ll y_0), \nonumber \\
            & R &= - \frac {(2n-1)(2j-1)\rho _y{}^2}
                          {i^2 \rho _x{}^2}\ \ \ (< 0).
\end{eqnarray}
\end{indention}
\end{indention}
Point A is a solution which was found by Masip\cite{Masip}.
At this point not only two terms in $g(x,y)$ cancel out
with each other in their leading order
but also the leading term in $f\,f_y$
of Eq.(19) cancels out $g\,g_y$.
In the expansion the ratio of the next-to-leading terms to
the leading ones is $O((y_0/x_0)^2)$.
In the case $j \geq 2$ Point C becomes a local minimum
only for $1+R \geq 0$ and is not a solution in the case $j=1$.
When $0 < \xi < 1$ Point A is the absolute minimum.
This condition on $\xi $ is translated as
\begin{equation}
      0 < c < i \left( \frac {2j}{i} \right)^{i/n}
             {\rm and} \ \ c \neq i \left( \frac {j}{i}
                          \right)^{i/n}.
\end{equation}
It is worth noting that under this condition the Point A
is the absolute minimum independent of the sign of $m_N{}^2$.
For illustration we show  the behavior of the scalarpotential
for the case $(n,i,m,j) = (6,3,2,1)$ in Fig.1.
In Fig.1 the vertical axis is taken as
\begin{equation}
        v = \left( 2M_C{}^4 \rho _x{}^2 x_0{}^2 \right)^{-1} V + 1
\end{equation}
and instead of $x$ and $y$ the horizontal axes are taken as
$\overline x = (x/x_0)^3$ and $\overline y = y/y_0 $
so that the point $(\overline x, \overline y) = (1, 1)$
becomes the absolute minimum (Point A).
In the case $(n,i,m,j) = (6,3,2,1)$ the condition (27) leads to
$0 < c < \sqrt 6$ and $c \neq \sqrt 3$.
In Fig.1 we put $a = b = c = 1$.
As seen in Fig.1, local minima (Points A and B) are located
at bottoms of very deep valleys.

\begin{center}
               {\large {\bf Fig.1}}
\end{center}

We are now in a position to evaluate the mass matrix for
$S, N$ and $\overline S, \overline N$
at the absolute minimum (Point A).
The components $(S-\overline S)/\sqrt 2$
and $(N-\overline N)/\sqrt 2$
are absorbed into massive vector superfields
due to the Higgs mechanism.
For the remaining components the mass matrix is of the form
\begin{equation}
\left(
\begin{array}{cc}
             O(1)            &     O( x_0/y_0 )    \\
             O( x_0/y_0 )    &     O( (x_0/y_0)^2 )
\end{array}
\right) m_{SUSY}
\end{equation}
with the basis $(S+\overline S)/\sqrt 2$
and $(N+\overline N)/\sqrt 2$.
Thus we obtain a large M-mass
\begin{equation}
          M_M = \frac {2(m-j)}{\sqrt {2n-1} \,\xi }
                 \left( c/i\right)^{n/i}
                  \sqrt  {-m_S{}^2} \left( {x_0}/{y_0} \right)^2,
\end{equation}
which is associated with the eigenstate
\begin{equation}
         \frac {1}{\sqrt 2}( N + \overline N )
               - \frac {i}{\sqrt 2 \,(m-j)}
                \left( {y_0}/{x_0} \right)( S + \overline S ).
\end{equation}
The enhancement factor $(x_0/y_0)^2$ depends on $n$ and $m$ as
\begin{equation}
         \left( {x_0}/{y_0} \right)^2 \sim
                   \left( {1}/{\rho _x} \right)^{(n-m)/(n-1)m}
\end{equation}
with $\rho _x{}^{-1} = 10^{15\sim 16}$.
Since the exponent $(n-m)/(n-1)m$ decreases with increasing $m$,
we take $m=2$ so as to get a sufficiently large M-mass $M_M$.
Then we have $j=1, n=2i$ and obtain
\begin{equation}
         \left( {x_0}/{y_0} \right)^2 = 10^{7 \sim 8}
                        \ \ \ \ \ {\rm for} \ \ n \geq 6.
\end{equation}
This means that the M-mass becomes
\begin{equation}
            M_M  = O\left( 10^{9 \sim 10} \gev \right)
\end{equation}
by taking $\sqrt {-m_S{}^2} = O(10^3\gev)$.
Consequently, a large M-mass can be induced from
the NR interactions of
$S, N$ and $\overline S, \overline N$
which are of the form
\begin{equation}
    W_{NR} = M_C{}^3 \lambda _1\,\biggl[
                 \left( \frac {S\overline S}{M_C{}^2} \right)^{n}
             + \frac {n}{2}
                    \left( \frac {N\overline N}
		{b^2\,M_C{}^2} \right)^2
             - 2c \left( \frac {S\overline S}{M_C{}^2} \right)^{n/2}
                    \left( \frac {N\overline N}{b^2\,M_C{}^2} \right)
                                      \biggr]
\end{equation}
with $0 < c < \sqrt{n}$ and $c \neq \sqrt{n/2}$.
For comparison we tabulate the orders of
$\langle S \rangle, \langle N \rangle $ and $M_M$
for several cases of the set $(n,i,m,j)$ in Table I.
As seen in this Table, unless $m=2$ and $j=1$,
$M_M$ attains to only at most $O(10^7\gev)$.
The case $m=2$ and $j=1$ is indispensable for solving
the solar neutrino problem.

\begin{center}
         {\large {\bf Table I}}
\end{center}

Untill now we consider the case (20).
In the other cases, for example,
\begin{eqnarray}
         \frac {n-i}{j} > \frac {n}{m} >
         \frac {i}{m-j} > 1, \nonumber \\
         1 < \frac {n-i}{j} < \frac {n}{m}
	< \frac {i}{m-j}, \nonumber
\end{eqnarray}
we have no interesting solutions in which
a large M-mass is derived from the minimization
of the scalarpotential.

In conclusion, we found that a large M-masss can be induced
from the NR interactions of
$S, N$ and $\overline S, \overline N$
in Calabi-Yau superstring models.
A pair of $S, N$ and $\overline S, \overline N$ chiral superfields
is needed to get this amazing result.
It is essential that two large intermediate energy scales
of symmetry breaking given by
$\langle S \rangle (\langle \overline S \rangle)$ and
$\langle N \rangle (\langle \overline N \rangle)$ emerge
as a consequence of the minimization of the scalarpotential.
This implies that the gauge symmetry should be rank-6
at the unification scale.
Furthermore, the special sets
$m=2, j=1, n=2i \geq 6$ in the NR interactions Eq.(8)
are necessary for realistic scenarios.
In fact, the M-mass becomes $O(10^{9 \sim 10}\gev)$ for these sets.
This large M-mass solves the solar neutrino promlem and also
is compatible with the cosmological bound for stable light neutrinos.
Special form of the NR terms suggests
that the superstring model possesses an appropriate
discrete symmetry coming from
distinctive structure of the compactified manifold.
The detailed study of the present
models will be presented elsewhere
\cite{Ours}.

\newpage

\vspace{2cm}

{\large {\bf Table Captions}}
\bigskip

{\bf Table I} \\
The energy scales of symmetry breaking $\langle S \rangle$
and $\langle N \rangle$ and a large Majorana-mass $M_M$ in
GeV unit for various cases of $(n,i,m,j)$.
Here we take $M_C=10^{18.5}$GeV and $\sqrt {-m_S{}^2}=10^3$GeV.

\vspace {5cm}

{\large {\bf Figure Captions}}
\bigskip

{\bf Fig. 1} \\
The structure of the scalarpotential
in the case $(n,i,m,j)=(6,3,2,1)$
with $a=b=c=1$.
The vertical axis is taken as the normalized scalarpotential
$v$ (see text).
The horizontal axes are $\overline x = (x/x_0)^3$ and
$\overline y = y/y_0$, where $x=\langle S \rangle /M_C$
and $y= \langle N \rangle /M_C$. \\
(a) The overview of the scalarpotential $v$.
The Point A (the absolute minimum) is located at
$(\overline x, \overline y) = (1,1)$
and the Point B is a local minimum. \\
(b) The comparison of values of the scalarpotential $v$
between Point A and Point B.
A solid (dashed) curve represents
the calculation of $v$ vs. $\overline x$
along the line $\overline x=\overline y \ (\overline y=0)$.

\newpage

\vspace{5cm}

\begin{center}
{\bf Table I} \\
\bigskip

\begin{tabular}{|cccc|cc|c|}\hline
  $n$   &   $i$   &   $m$   &   $j$
                & $\langle S \rangle$ (GeV)
                      &   $\langle N \rangle$ (GeV)
                            &   $M_M$ (GeV)         \\ \hline
\hline
 4  &  2  &  2  &  1  &  $10^{15.9}$
&  $10^{13.1}$  &  $10^{8.1}$ \\
 6  &  3  &  2  &  1  &  $10^{16.9}$
&  $10^{13.5}$  &  $10^{8.8}$ \\
 8  &  4  &  2  &  1  &  $10^{17.4}$
&  $10^{13.6}$  &  $10^{9.1}$ \\
10  &  5  &  2  &  1  &  $10^{17.6}$
&  $10^{13.7}$  &  $10^{9.2}$ \\
12  &  6  &  2  &  1  &  $10^{17.8}$
&  $10^{13.7}$  &  $10^{9.2}$ \\
20  & 10  &  2  &  1  &  $10^{18.1}$
&  $10^{13.7}$  &  $10^{9.2}$ \\ \hline

 6  &  4  &  3  &  1  &  $10^{16.7}$
&  $10^{14.7}$  &  $10^{6.6}$ \\
 9  &  6  &  3  &  1  &  $10^{17.5}$
&  $10^{15.3}$  &  $10^{6.4}$ \\
12  &  8  &  3  &  1  &  $10^{17.8}$
&  $10^{15.4}$  &  $10^{6.6}$ \\ \hline

 6  &  2  &  3  &  2  &  $10^{16.9}$
&  $10^{15.2}$  &  $10^{5.4}$ \\
 9  &  3  &  3  &  2  &  $10^{17.5}$
&  $10^{15.2}$  &  $10^{5.8}$ \\
12  &  4  &  3  &  2  &  $10^{17.8}$
&  $10^{15.3}$  &  $10^{5.9}$ \\ \hline
\end{tabular}
\end{center}


\newpage

\begin{thebibliography}{1}

\bibitem{gaugesym}
S.Cecotti, J.P.Derendinger,
	S.Ferrara, L.Girardello and M.Roncadelli,
       Phys. Lett. {\bf 156B} (1985) 318. \\
J.D.Breit, B.A.Ovrut and G.C.Segre,
	Phys. Lett. {\bf 158B} (1985) 33. \\
M.Dine, V.Kaplunovsky, M.Mangano, C.Nappi and N.Seiberg,
         Nucl. Phys. {\bf B259} (1985) 549. \\
T.Matsuoka and D.Suematsu, Nucl. Phys. {\bf B274} (1986) 106;
          Prog. Theor. Phys. {\bf 76} (1986) 886.

\bibitem{seesaw}
M.Gell-Mann, P.Ramond and S.Slansky,
	in Supergravity, eds. D.Freedman
    {\it et al.} (North-Holland, Amsterdam, 1979). \\
T.Yanagida, KEK lectures, eds. O.Sawada {\it et al.} (1980)912;
      Phys. Rev. {\bf D23} (1981) 196. \\
R.Mohapatra and S.Senjanovic, Phys. Rev. Lett. {\bf 44} (1980) 912.

\bibitem{neutrino}
S.P.Mikheyev and A.Y.Smirnov, Yad. Fiz. {\bf 42} (1985) 1441;
             Nuov. Cim. {\bf 9C} (1986) 17. \\
L.Wolfenstein, Phys. Rev. {\bf D17} (1978) 2369;
               Phys. Rev. {\bf D20} (1979) 2634. \\
S.A.Bludman, D.Kennedy and P.Langacker,
Phys. Rev. {\bf D45} (1992) 1810;
           Nucl. Phys. {\bf B374} (1992) 373.

\bibitem{Cowsik}
R.Cowsik and J.McClelland,
	Phys. Rev. Lett. {\bf 29} (1972) 669.

\bibitem{discrete}
C.A.Lutkin and G.G.Ross, Phys. Lett. {\bf 214B} (1988) 357. \\
C.Hattori, M.Matsuda, T.Matsuoka and H.Mino,
             Prog. Theor. Phys. {\bf 82} (1989) 599.

\bibitem{Masip}
M.Masip, Phys. Rev. {\bf D46} (1992) 3601.

\bibitem{Zoglin}
P.Zoglin, Phys. Lett. {\bf 228B} (1989) 47.

\bibitem{Ours}
N.Haba, C.Hattori, M.Matsuda
T.Matsuoka and D.Mochinaga, in preparation.


\end{thebibliography}
\end{document}